\documentclass[aps,prb,twocolumn,superscriptaddress,showpacs]{revtex4}
\usepackage{amsmath,amssymb}
\usepackage{graphicx}
\usepackage{psfrag}

\newcommand{\ek}{\epsilon_\mathbf{k}}
\newcommand{\ed}{\epsilon_d}

\newcommand{\integral}[3]{\int\limits_{#2}^{#3}\!d#1}          
\newcommand{\gf}[3]{#1_{#2}^{#3}}                              
\newcommand{\crop}[2]{#1_{#2}^\dagger}                         
\newcommand{\deop}[2]{#1_{#2}^{\phantom{\dagger}}}             
\renewcommand{\Re}{\mathsf{Re}}                                
\renewcommand{\Im}{\mathsf{Im}}                                

\begin{document}

\title{$\pi$-Junction behavior and Andreev bound states\\
 in Kondo quantum dots with superconducting leads}

\author{Gabriel Sellier}
\affiliation{EP VI, Institut f\"ur Physik, 
Universit\"at Augsburg, 86135 Augsburg, Germany}
\email{thilo.kopp@physik.uni-augsburg.de}
\affiliation{TKM, Universit\"at Karlsruhe, Postfach 6980, 
76128 Karlsruhe, Germany}

\author{Thilo Kopp}
\affiliation{EP VI, Institut f\"ur Physik, 
Universit\"at Augsburg, 86135 Augsburg, Germany}

\author{Johann Kroha}
\affiliation{Physikalisches Institut, Universit\"at Bonn, 
Nussallee 12, 53115 Bonn, Germany}

\author{Yuri S.~Barash}
\affiliation{Institute of Solid State Physics,
Russian Academy of Sciences, Chernogolovka, 142432 Moscow, Russia}

\date{\today}

\begin{abstract}
We investigate the temperature- and coupling-dependent transport through  
Kondo dot contacts with symmetric superconducting $s$-wave leads.  
For finite temperature $T$ we use a superconducting extension 
of a selfconsistent auxiliary boson scheme, termed SNCA, while at
$T=0$ a perturbative renormalization group treatment is applied.   
The finite-temperature phase diagram for the $0$--$\pi$ transition 
of the Josephson current in the junction is established and related
to the phase-dependent position of the subgap Kondo resonance 
with respect to the Fermi energy. The conductance of the contact is
evaluated in the zero-bias limit. It approaches zero in the
low-temperature regime, however, at finite $T$ its characteristics are 
changed through the coupling- and temperature-dependent $0$--$\pi$ 
transition.
\end{abstract}


\maketitle

\section{Introduction}

The physics of charge and spin transport through Kondo quantum dots
is paradigmatic for interface problems with strong
correlations mediated by the contact. Novel effects, taking
place in junctions with quantum dots between normal metal leads,
have been intensively studied for a long 
period.\cite{Appelbaum66,Anderson66,Glazman88,Lee88,Aleiner02}
The Josephson current through a localized spin state was first
considered by Shiba and Soda. \cite{Shiba69} Later on Glazman and Matveev
investigated more thoroughly the supercurrent through a
single resonant state, as well as through a distribution of such
impurity states~\cite{Glazman89}. 
Eventually, a Kondo quantum dot, which is coupled to a normal and to a
superconducting lead, is a further notable system which allows to approach
the interplay between Kondo effect and Andreev reflections.~\cite{Schwab99,Cuevas01}
The progress in the miniaturization of
electronic devices now makes the investigation of electronic
transport through a single Kondo impurity technically feasible.
To date, several groups have reported on transport measurements of 
such nanoscale 
devices \cite{Scheer97,Scheer98,Sasaki00,Ralph02,Schoenenberger02,
Agrait03,Schoenenberger04}. Besides possible applications
as, for example, the study of nonlocal spin-entangled pairs~\cite{Recher01},
these quantum dot contacts are fascinating on fundamental grounds,
because they are the most elementary realization of 
a ``strongly correlated contact''.

From the theoretical side it has been well apprehended that a
phase-sensitive subgap state is formed, which is to be interpreted
as a Kondo resonance, if the Kondo scale $T_K$ is larger than the
gap $\Delta$ of the superconducting leads, $T_K/\Delta \gg 1$.
Andreev scattering processes induce a dependence of the 
subgap-state energy on the phase difference $\phi$
of the superconductors in the leads.\cite{Clerk00b} 
The interference of Andreev scattering with Kondo-type spinflip
processes leads to a non-trivial behavior of the
Josephson current--phase relation  $I_s(\phi )$ in Kondo quantum dot
junctions. A transition from a $0$-junction to a $\pi$-junction arises 
on account of the distinct nature of the  spin ground states in the
strong and in the weak coupling regime. 
In the strong coupling limit, $T_K/\Delta \gg 1$, the ground state 
is a spin singlet due to the Kondo screening of the impurity spin, 
and the Coulomb blockade is lifted by the formation of the Kondo resonance.
In this case, coherent Cooper pair transmission occurs 
without affecting the spin of the electrons in the pair ($0$-junction). 
In the opposite limit of weak coupling, $T_K/\Delta \ll 1$,
the Kondo screening is suppressed for temperatures $T$ well below the
critical temperature $T_c$ of the superconductors as the Cooper
pairs in the bulk cannot be broken for $s$-wave pairing symmetry.
Then the ground state is a Kramers degenerate spin doublet, 
and a single subgap resonance of width $\sim T_K$ in the 
impurity spectrum is formed, split off from the continuum spectrum. 
In this regime, retarded, coherent pair transmission is still 
possible, but for energetic reasons (no double occupancy of the
dot) the temporal sequence of the transmitted electrons with opposite spin 
is reversed, leading to a $\pi$-shift in the current--phase relation
($\pi$--junction),\cite{Spivak91} see Ref.\ \onlinecite{Rozhkov00} for a
more detailed discussion. Since for weak coupling the 
subgap state forms below  and for strong coupling it moves above the
Fermi energy,\cite{Clerk00b,Kirchner04} the current--phase relation $I_s(\phi)$
is also related to the position of the resonance. This behavior of a
Kondo quantum dot should be contrasted to the case when the impurity 
state is not a dynamical quantity and its magnetic moment is fixed. 
The latter case is analogous to junctions with  ferromagnetic interlayers, 
where Andreev subgap states are generated both below and above the Fermi
energy, being split with respect to their spin
polarization.\cite{Fogelstroem00,Barash02,Andersson02,Bruder01} 

Whereas the supercurrent through a Kondo correlated junction has been
investigated successfully within several approaches, the
conductance of the contact is much more difficult to study, as it
involves the quasiparticle current. 
It is essential to distinguish between {\it Kondo point
contacts} and {\it Kondo quantum dot devices}.
A Kondo impurity in a point contact or orifice introduces an 
additional scattering channel and tends to reduce the transmission
similar to the Kondo effect in bulk metals. In contrast, for a Kondo
quantum dot device, the quantum dot provides the only transmission
channel. For temperatures above $T_K$ and a quantum dot energy
level well below the Fermi energy, the transmission channel is
``almost closed'' as the charge tunneling is suppressed by the
Coulomb blockade and because the level is off resonance. Through the formation 
of the Kondo resonance at temperatures below $T_K$, on-resonance
tunneling enhances the transmission up to the quantum
limit.~\cite{Glazman88,Lee88,Pustilnik04}
While this inverse relation between the two junction types is rather
obvious for normal conducting leads, it has more profound consequences 
in the case of superconductors. 
In the present paper we study the intrinsic conductance which
characterizes the quasiparticle current through Kondo dots between
two superconductors. The quasiparticle current is not related in
a simple way to the supercurrent and the question arises, if the
$0$--$\pi$ transition may already manifest itself in the zero bias
conductance of the Kondo dot contact. Choi {\it et al.}~\cite{Choi04}
also investigate a Kondo quantum dot with superconducting leads 
to calculate the Josephson current at $T=0$. However, to determine the 
low-$T$ conductance through the $0$--$\pi$ transition, they consider 
a Kondo quantum dot with an additional resistive shunting  
(resistively shunted superconducting
junction, RSCJ) in the overdamped regime and 
compare the crossover for $T_K\approx 0.5\Delta$ with the measured 
conductance of gated carbon nanotube quantum dots coupled to superconducting
Au/Al leads~\cite{Buitelaar02}.  Although the RSCJ modelling 
may well apply to the considered experiments, it actually does not refer
to a Kondo quantum dot as defined above. It does not analyze the
quasiparticle current through the Kondo impurity 
but rather the phase slips of the supercurrent. The respective 
conductance $G_S$ in the RSCJ model grows exponentially with the 
inverse temperature $G_S/G_N \sim \exp(\hbar I_s/eT)$ where $G_N$ is 
the conductance in the normal state.
 

In this article we address two related subjects which are relevant
for Kondo dots between two superconducting leads. On the one hand,
a finite-temperature phase diagram for the $\pi$-junction behavior
has not yet been presented. It allows to identify the coupling
strength at which the $0$--$\pi$ transition sets in, but it also
renders the regime where the transition may be observed by varying
the temperature at fixed coupling strength. On the other hand, we
calculate the intrinsic conductance of the Kondo dot contact ---
a quantity which is not related to the phase difference as 
directly as the supercurrent. Nevertheless, as both
quantities, the supercurrent and the quasiparticle current, are
controlled by the position of the subgap resonance, they both
display a transition in their dependence on temperature and
coupling strength, as we will discuss in this paper. For this
investigation we will always consider the generic case of a
symmetric junction and  $s$-wave pairing symmetry in the leads.

In Sec.~\ref{sec:technique} we briefly introduce the technique used
for the Kondo correlated junctions,
an extension of the non-crossing approximation (NCA)  
to the superconducting state (SNCA).
Some explicit details about the derivation of the SNCA, 
its evaluation and regime of validity, as well as the calculation of the
supercurrent and the conductance are deferred to the appendices.
Sec.~\ref{sec:supercurrent} expands the discussion of the $0$--$\pi$
transition in certain aspects beyond what has been presented in the
literature on this topic. 
Specifically, we focus on the temperature dependence of
the transition and introduce a phase diagram.
In Sec.~\ref{sec:conductance} we address
the temperature- and coupling-dependence of the intrinsic
conductance of the Kondo dot contact.

\section{Kondo Impurity between two Superconductors}
\label{sec:technique}

\begin{figure}[t]
  \includegraphics[width=4.5cm]{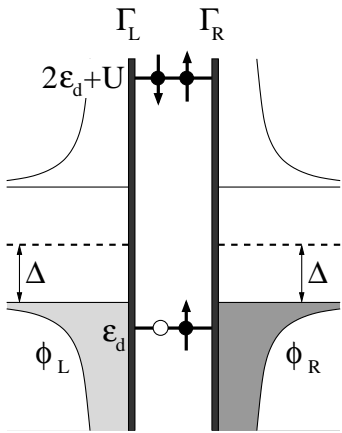}
  \caption{\label{fig:squantumdot}Quantum dot coupled
           to two superconductors. $\Gamma_L$ and $\Gamma_R$ denote
           the effective couplings to the left and right
           lead, $\phi_L$ and $\phi_R$ label the phases of the
           corresponding superconducting order parameter. The BCS
           gap $\Delta$ is assummed to be equal in both superconductors.
           In the model, defined by Eq.~(\ref{eq:bcsanderson}), 
           the local Coulomb repulsion $U$ is set to infinity.}
\end{figure}

The system of conduction electrons in the left and right lead
interacting with a single-channel magnetic impurity or quantum dot
is modeled by an infinite-$U$ Anderson Hamiltonian. 
The $s$-wave superconducting state in the
reservoirs is treated within standard BCS mean-field theory. The
complete Hamiltonian then takes the form (see Fig.~\ref{fig:squantumdot}
for a graphical layout of the contact)
\begin{align}
   \label{eq:bcsanderson}
   &H=H_0+H_{BCS}+H_{QD}+\lambda\,Q\,\\
   \intertext{with}
   &H_{0}  =\sum\limits_{\mathbf{k}\sigma}
                   \ek\crop{c}{\mathbf{k}\sigma a}\deop{c}{\mathbf{k}\sigma a}\,,\notag\\
   &H_{BCS}=-\sum\limits_{\mathbf{k}a}\Delta_a\left(
                   \crop{c}{\mathbf{k}\uparrow a}\crop{c}{\mathbf{-k}\downarrow a}+
                   \mathsf{h.c.}\right)\,,\notag\\
   &H_{QD}=\sum\limits_{\sigma}
                   \ed\crop{f}{\sigma}\deop{f}{\sigma}
                   +\sum\limits_{\mathbf{k}\sigma}V_a\left(
                   \crop{c}{\mathbf{k}\sigma a}\crop{b}{}\deop{f}{\sigma}+
                   \mathsf{h.c.}\right)\,.\notag
\end{align}
Here we have adopted a slave-boson representation \cite{Barnes79}
for the dot states,
where the local creation and annihilation operators
for an electron in the dot ($d$-) orbital with spin $\sigma$ and 
energy $\epsilon _d$ are decomposed as, for example,
$\crop{d}{\sigma} =\crop{f}{\sigma} b$. The operators $\crop{f}{\sigma}$
and $\crop{b}{}$ create a singly occupied or an empty occupied impurity state,
whenever an electron hops onto or off the dot, respectively, and
obey the canonical fermion and boson commutation relations.
Their dynamics are restricted to the physical Hilbert space by the
operator constraint
\begin{align}
   \label{eq:constraint}
   Q=\sum\limits_{\sigma}\crop{f}{\sigma}\deop{f}{\sigma}
     +\crop{b}{}\deop{b}{}=1\, ,
\end{align}
which will be enforced exactly by taking the limit of the parameter
$\lambda \to \infty$ (see Appendix A).\cite{Coleman84}
Moreover, $\crop{c}{\mathbf{k}\sigma a}$ creates a conduction electron
in the left (L) or right (R) superconductor, $a=L,R$. The
hybridization of these electronic states in the leads with the quantum
dot state is parameterized by $V_a$. For convenience, we introduce the
effective couplings $\Gamma_a=\pi N_0 V_{a}^2$ and 
$\Gamma = \Gamma_L +\Gamma_R$, where
$N_0$ is the density of states at the Fermi energy in the normal
conducting state. 
%

The BCS part of the Hamiltonian can be easily solved. 
The normal and the anomalous 
local advanced/retarded conduction electron Green's functions 
are defined as
$\gf{G}{a}{A/R}(t)=\pm\sum_\mathbf{k}i\theta(\mp t)
    \langle\{\deop{c}{\mathbf{k}\sigma a}(t),\crop{c}{\mathbf{k}\sigma a}(0)\}
\rangle$
and
$\gf{F}{a}{A/R}(t)=\pm\sum_\mathbf{k}i\theta(\mp t)
    \langle\{ \deop{c}{\mathbf{k}\uparrow a}(t),
              \deop{c}{-\mathbf{k}\downarrow a}(0)\} \rangle$,
respectively.
The gap equations defining the order parameter $\Delta _a$ in the two
superconductors are given by
$\Delta_a=V_{BCS}^{\phantom{\dagger}}\sum_{\mathbf{k}}
\langle\deop{c}{\mathbf{-k}\downarrow a}\deop{c}{\mathbf{k}\uparrow a}\rangle.$
In the subsequent consideration the amplitude of $\Delta_a$ is
assumed to be equal on both sides, i.e.
\begin{align}
  \label{eq:orderparameter}
  \Delta_{a}=|\Delta|e^{i\phi_{a}}\;.
\end{align}
For the local conduction electron density of states per spin
and the corresponding anomalous contribution one obtains,
\begin{subequations}
\label{eq:localDOS}
\begin{alignat}{2}
  &\rho_{a}(\epsilon)=
         \frac{\gf{G}{a}{A}(\epsilon)-\gf{G}{a}{R}(\epsilon)}{2\pi i N_0}
        &=+\Re\frac{|\epsilon|}{\sqrt{\epsilon^2-|\Delta_a|^2}}\,,\\
  &g_a(\epsilon)=
         \frac{\gf{F}{a}{A}(\epsilon)-\gf{F}{a}{R}(\epsilon)}{2\pi i N_0}
        &=-\Re\frac{\mathrm{sign}(\epsilon)\Delta_a}{\sqrt{\epsilon^2-|\Delta_a|^2}}\,,
\end{alignat}
\end{subequations}
where both spectral functions have been normalized to $N_0$.

\begin{figure}[t]
  \begin{center}
    $\Phi\;\;=\;\;$
    \begin{minipage}[c]{1.43cm}
      \includegraphics[width=1.43cm]{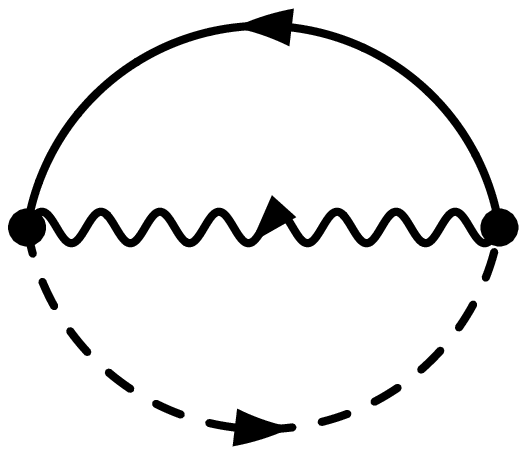}
    \end{minipage}
    $\;\;+\;\;$
    \begin{minipage}[c]{1.57cm}
      \includegraphics[width=1.57cm]{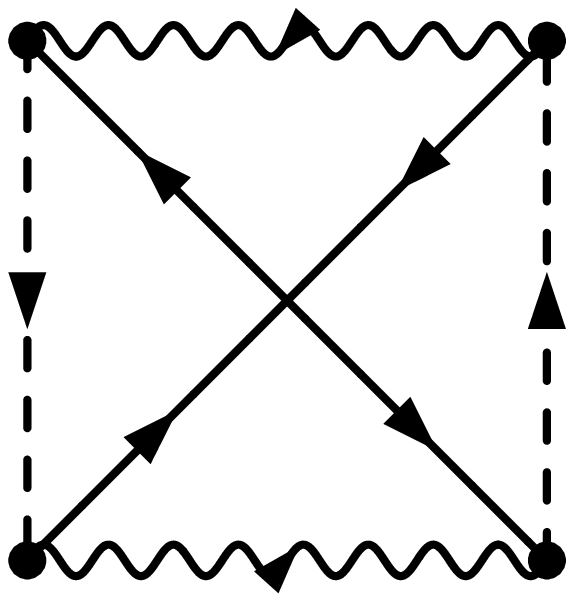}
    \end{minipage}
    \caption{Generating functional for the extension of the NCA to
             the broken-symmetry state (SNCA). The solid, wavy and dashed
             lines represent the conduction electron, slave boson and 
             pseudofermion propagators, respectively.}
    \label{fig:GenFuncSNCA}
  \end{center}
\end{figure}

For the greater part of this paper the Kondo dynamics of the
quantum dot at finite temperatures will be described within 
a selfconsistent approach, where the local gauge symmetry
on the dot is preserved by means of conserving approximations,
derived from a Luttinger-Ward functional \cite{Kadanoff61}. 
We will use a generalization of the well-known non-crossing approximation 
(NCA)\cite{Keiter71,Kuramoto83} 
for superconducting leads, the ``superconducting NCA'' (SNCA), to
include retarded Cooper pair tunneling. The 
Luttinger-Ward generating functional $\Phi$ for the SNCA is
depicted in Fig.~\ref{fig:GenFuncSNCA}. The leading term of 
$O(\Gamma )$ in a selfconsistent expansion corresponds to the NCA
(first diagram in Fig.\ \ref{fig:GenFuncSNCA}).
For normal conducting leads, the NCA is known to give a satisfactory,
quantitative description of the spectral features in the case of 
infinite $U$,\cite{Sakai88,Pruschke89,Holm89,Costi96,Haule01} 
in the absence of magnetic field,\cite{Kirchner02,Kroha05} 
and for temperatures down to $T\approx 0.1\ T_K$.\cite{Costi96}. 
However, in the case of superconducting leads the NCA 
completely neglects Andreev scattering contributions
(Cooper pair tunneling through the dot), 
which is crucial for the Josephson current
and which will also induce significant renormalizations of the 
normal quasiparticle current, as seen below.   
Therefore, the NCA is extended to include the
next-to-leading term of order $O(\Gamma ^2)$ 
(second diagram in Fig.\ \ref{fig:GenFuncSNCA}), which contains
two anomalous lead Green's functions, constituting the  
SNCA.\cite{Clerk00a} Similar, but simplified methods, employing an
elastic scattering approximation,
have also been used by Bickers and Zwicknagl\cite{Bickers87} and
by Borkowski and Hirschfeld\cite{Borkowski94}.
A detailed discussion of the SNCA is deferred to Appendix A.
It will be seen that the SNCA describes, to leading selfconsistent 
order, the coherent transmission of Cooper pairs via the formation
of {\it retarded} Cooper pairs on the dot even though the existence
of {\it equal-time} Cooper pairs on the dot is prohibited by the
local Coulomb repulsion. Superconducting Kondo dot junctions have 
recently been considered also within a mean field approach to 
the dot dynamics,\cite{Avishai01,Avishai03} 
which tends to overestimate the Cooper pair correlations 
on the dot due to the assumption of static rather than retarded pairs 
on the dot.
\par
From the generating functional in Fig.~\ref{fig:GenFuncSNCA}
a set of coupled integral equations for the pseudoparticle
selfenergies can be derived as well as an expression for
the local spectral function of the quantum dot and its
corresponding anomalous part. These equations are solved
numerically. The explicit expressions are discussed 
in Appendix~\ref{ap:SNCAformulae}.\par
To calculate the Josephson current we use the formula first presented by
Clerk and Ambegaokar\cite{Clerk00b}
\begin{align}
\label{eq:critical_current}
  I_s(\phi)=\frac{2e}{h}\frac{\Gamma}{\pi N_0}\sin\phi&\integral{\omega}{}{}\,f(\omega)\notag\\
            &\times\Im\left[\gf{\bar{\mathcal{F}}}{d}{R\, \dagger}(\omega)
                     \gf{\bar{F}}{}{R}(\omega)\right]\,,
\end{align}
which is rederived in Appendix~\ref{ap:joscurrent}. The quantities in
this current relation are defined as follows: $\phi=\phi_L-\phi_R$ denotes 
the phase difference between left and right lead, 
$f(\omega)$ is the Fermi function and, for convenience, 
we extracted the explicit phase
dependence from the off-diagonal Green's functions,
\begin{align*}
   &\gf{\mathcal{F}}{d}{R}(\omega)=\cos\left(\frac{\phi}{2}\right)
                                          \gf{\bar{\mathcal{F}}}{d}{R}(\omega)
                                          \\
   &\gf{F}{a}{R}(\omega)=e^{i\phi_a}\gf{\bar{F}}{}{R}(\omega)\,,
\end{align*}
where $\gf{\mathcal{F}}{d}{R\, \dagger}(\omega)$ and 
$\gf{F}{a}{R}(\omega)$ are the anomalous parts of
the Green's function of the impurity $d$-level and of the 
conduction-electron Green's function in lead~$a$, respectively
(cf.~Appendix~\ref{ap:joscurrent}).

The zero bias conductance $G=dI/dV|_{V=0}$ is calculated from the
quasiparticle current in the limit of small bias\cite{Meir92},
\begin{equation}
\label{eq:conductance}
   G=-2\frac{e^2}{h}\Gamma\int\! d\omega\;\frac{\partial
   f(\omega)}{\partial\omega}\; \rho(\omega)\;
    \Im\gf{\mathcal{G}}{d}{A}(\omega)  \,,
\end{equation}
with $\gf{\mathcal{G}}{d}{A}(\omega)$ the normal part of the
impurity Green's function. 
As we consider a symmetric coupling to the two leads with equal
spectral densities ($\Gamma_L = \Gamma_R
\equiv \pi N_0 V_{L,R}^2$, $\Gamma \equiv \Gamma_L + \Gamma_R =
2\Gamma_{L,R}$, and $\rho_{L}(\omega)=\rho_{R}(\omega)\equiv\rho(\omega)$), 
all contributions with anomalous as well as with Keldysh Green's functions
vanish in Eq.~(\ref{eq:conductance}).
\section{Supercurrent}
\label{sec:supercurrent}
\begin{figure}[t]
  \begin{center}
      \includegraphics[width=7.9cm]{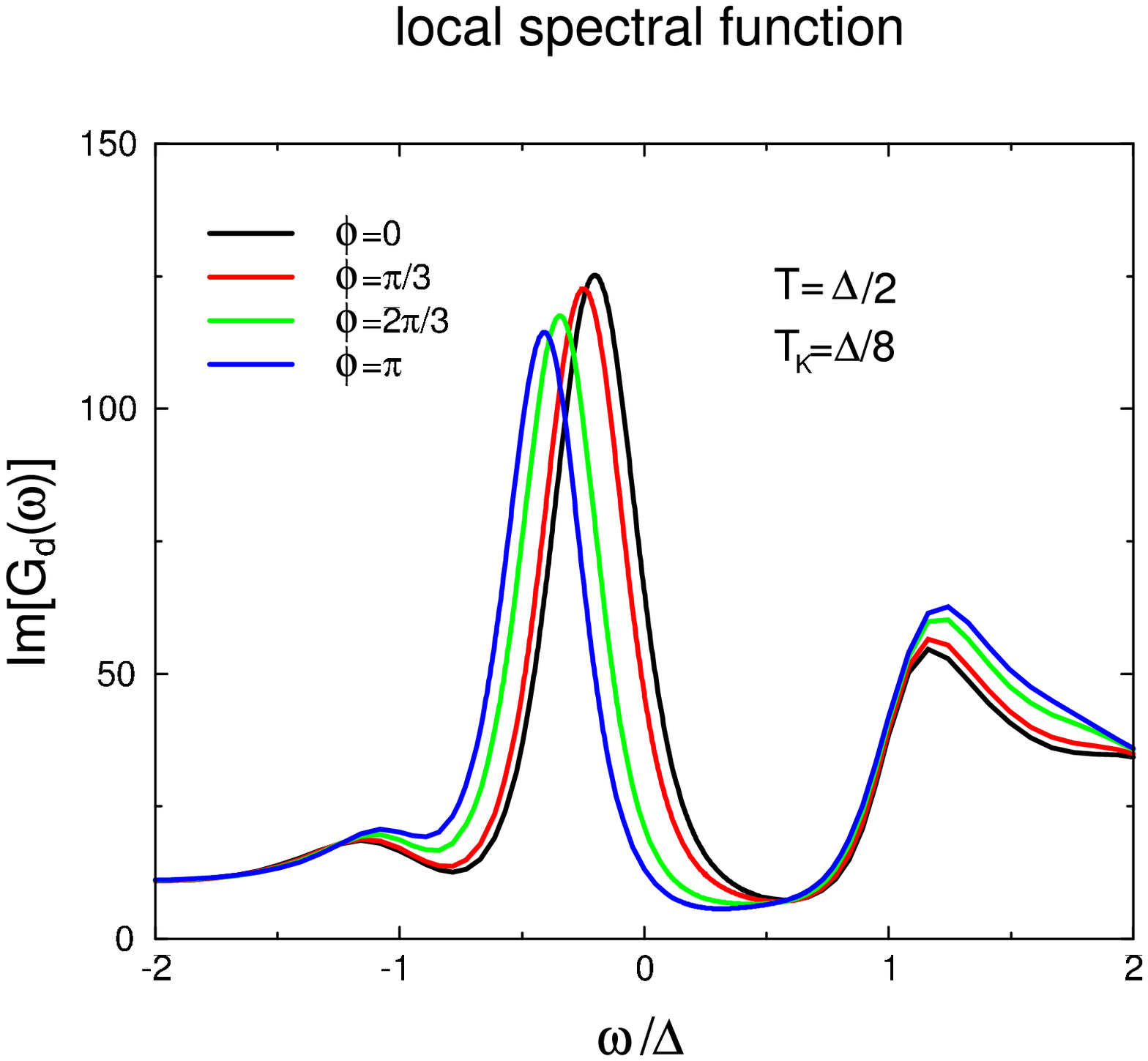}
\vskip 0.3cm
      \includegraphics[width=7.9cm]{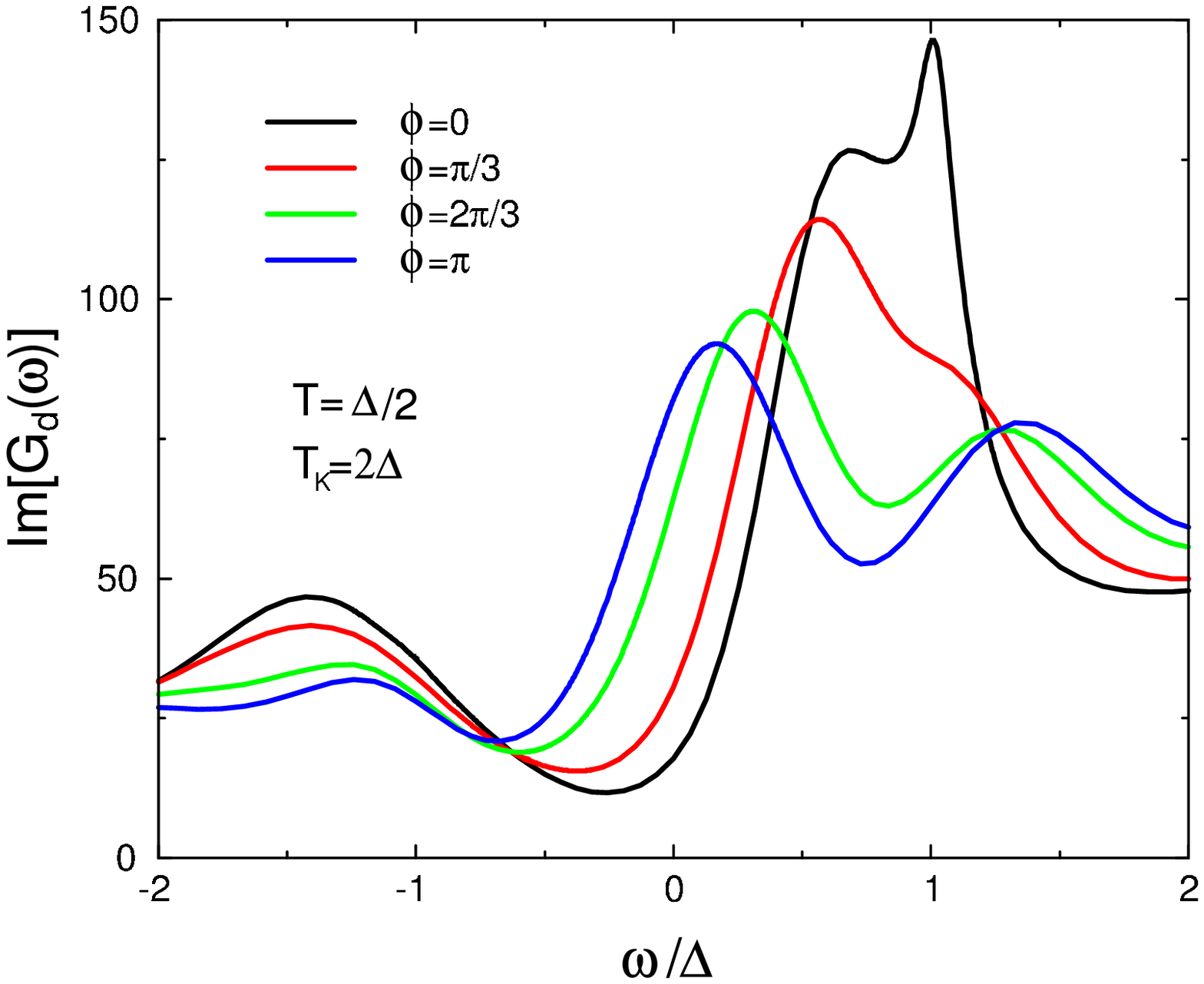}
\caption{The spectral function of the impurity $d$-level for weak
coupling (upper panel, $T_K/\Delta=0.125$) and strong coupling
(lower panel, $T_K/\Delta=2.0$). The leads are in the superconducting
state at $T=0.5\Delta$. The spectral functions are
sensitive to the phase difference $\phi=\phi_L -\phi_R$ between left
and right lead.
}
    \label{fig:spdens}
  \end{center}
\end{figure}

We now investigate the current-phase relation $I_s(\phi)$
in the parameter space which is controlled by temperature $T$ and
coupling strength $T_K/\Delta$. 
This analysis focuses naturally on a calculation at finite $T$,
for which regime the NCA yields quantitatively well controlled results\cite{Costi96}
in the absence of magnetic field, and so is expected to do its superconducting 
extension, the SNCA. We will compare the $T\to 0$ extrapolation of these calculations 
with our perturbative renormalization group (RG) analysis of the same model at $T=0$
as well as with exact numerical renormalization group (NRG) 
calculations by Choi {\it et al.}\cite{Choi04} 
at $T=0$  for the symmetrical Anderson model.  

\begin{figure}[t]
\vskip 0.5cm
  \begin{center}
      \includegraphics[width=8.8cm]{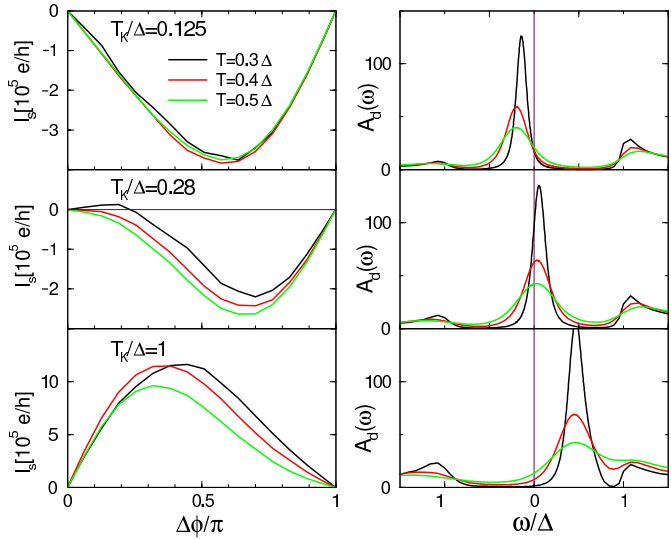}
    \caption{Phase dependence of the Josephson current for weak,
      intermediate and strong coupling values of $T_K/\Delta$
      (left column of panels).
      In the right panel, the spectral function of the impurity
      $d$-level is presented for each coupling strength and temperature.}
    \label{fig:Ic}
  \end{center}
\end{figure}

It has been elaborated by Clerk and
Ambegaokar~\cite{Clerk00b} that strong and weak coupling regimes are to be
distinguished by the position of the subgap resonance: the resonance moves
through the Fermi energy from below when the coupling is increased, a
behavior to be associated with a transition from a $\pi$- to a
0-junction type. In fact, we confirm this behavior in
Figs.~\ref{fig:spdens} and \ref{fig:Ic}.
However, the resonance is wider than observed by Clerk and Ambegaokar:
For strong coupling $T_K/\Delta >1$, the subgap resonance as well as the
features at the gap edges in the $d$-electron 
spectrum are of the order of $T_K > \Delta$, 
as Cooper pairs are broken in order to screen the
impurity spin in this regime. The gap edges are less pronounced 
(see Fig.~\ref{fig:spdens}, lower panel), and a Fano-like
interference between the continuum states and the subgap mode is
evident. 

\begin{figure}[t]
  \begin{center}
          \includegraphics[width=9.0cm]{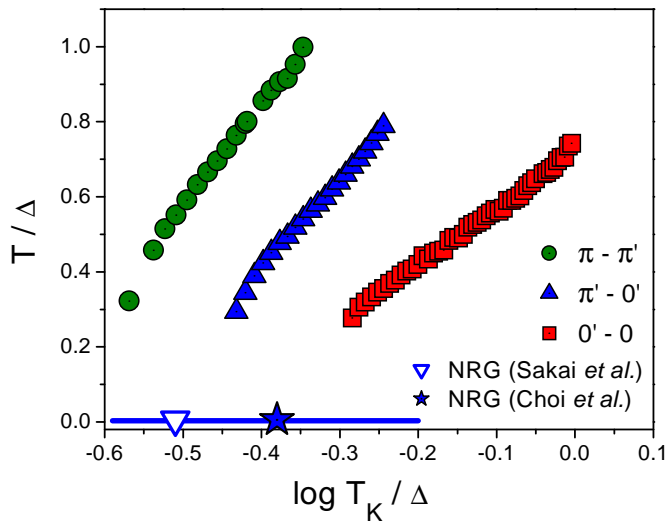}
    \caption{
            Phase diagram for the $0-\pi$ transitions. As discussed in the text,
            the lower right area corresponds to the 0-phase and the upper left 
            area to the $\pi$-phase.
            The SNCA data points refer to the $\pi-\pi'$ transition
            (circles), the $\pi'-0'$ transition (triangles), and the $0'-0$ transition
            (squares). As seen from these curves, the respective transition points as a
            function of $T$ scale roughly with ${\rm log} T_K$.
            The open inverse triangle presents the
            transition point from a spin doublet to a
            singlet state of a Kondo impurity in a bulk superconductor
            within the NRG evaluation of Satori {\it et al.}~\cite{Satori92}.
            The star on the horizontal axis is
            approximately the transition point in the NRG analysis for the
            symmetric Anderson model of the quantum dot contact 
            (Choi {\it et al.}~\cite{Choi04}).
            The fat line at the horizontal axis is the regime where the
            perturbative RG analysis suggests the  $\pi'-0'$ transition at
            zero temperature.
}
    \label{fig:phasediagram}
  \end{center}
\end{figure}

The current--phase relation traverses three scenarios or transitions,
as the coupling parameter $T_K/\Delta$ is raised from weak to strong coupling
(left column of Fig.~\ref{fig:Ic}).
These scenarios are related to the fact that the Josephson current states of 
a superconducting junction are equilibrium states and are thus determined by the minima 
of the free energy. One may identify a succession of four current-carrying
equilibrium states: 
$0$-junction: single global minimum for $\phi =0$;
$0'$-junction: global minimum for $\phi =0$ and local minimum for $\phi=\pi$;
$\pi '$-junction: local minimum for $\phi =0$ and global minimum for $\phi=\pi$;
$\pi$-junction: single global minimum for $\phi =\pi$.
The succession of the corresponding transitions has been discussed in the
literature~\cite{Rozhkov99,Siano04}. 
\begin{enumerate}
\item
{\it Weak coupling.}
In the case $T_K/\Delta=1/8$ all three curves $I_s(\phi)$
correspond to temperatures above $T_K$. The curves are nearly
identical within  numerical resolution. We observe a $\pi$-junction
behavior where the first harmonic, -$\sin\phi$, dominates. In SNCA
we cannot approach the low temperature limit where $T$ is
well below $T_K$. The latter has been investigated by Choi {\it et
al.}~\cite{Choi04} who indeed find a sinusoidal behavior for the
zero temperature limit in the weak coupling regime.
\item
{\it Intermediate coupling.}
Here, higher harmonics become important (as in the
second row of Fig.~\ref{fig:Ic} for  $T_K/\Delta=0.28$).
The derivative of the current at $\phi=0$ changes sign with
the temperature somewhere close to   $T=0.4\Delta$ (middle left
panel), which corresponds approximately to temperature and coupling
where the spectral function crosses the Fermi energy.
Choi {\it et al.}~\cite{Choi04} observe a
discontinuous behavior in the current-phase relation for the
intermediate coupling regime at zero temperature however this
discontinuity is smoothed for finite temperature~\cite{Choi05}.
The distinction between the
$\pi-\pi'$, the $\pi'-0'$ and the $0'-0$ transitions, which classify the
appearance and vanishing of the two minima of the free energy  
as mentioned above, is made by the
characteristics of the current-phase relation:
The sign reversal of the slope of $I_s(\phi )$ at vanishing $\phi$ 
in the intermediate coupling regime signifies the  $\pi'-\pi$ transition.
\item
{\it Strong coupling.}
The lower left panel in Fig.~\ref{fig:Ic} shows the current-phase relation
for  $\Delta/T_K=1$. For this value of the coupling we are already
in the strong coupling regime in the sense that the subgap resonance
is clearly above the Fermi energy and the  supercurrent is positive in the
considered phase interval (0-junction behavior). The temperature of all curves is below
$T_K$. The curve is more sinusoidal for the lowest temperature whereas it
develops a flatter region for $\phi$ close to $\pi$ for the higher temperatures.
This is a precursor to the $0-\pi$ transition and we will see below
that the $\pi$-junction behavior may be recovered for higher
temperature if $T_K/\Delta$ is not too large.
\end{enumerate}
\noindent The phase diagram is now derived from the analysis of the
extrema in the free energy: (i) the  $\pi$--$\pi'$ transition takes place when 
the maximum at phase $\phi=0$ turns into a local minimum of the free
energy which is equivalent to the sign change of the slope of the current-phase 
relation for vanishing
$\phi$ (circles in Fig.~\ref{fig:phasediagram}); (ii) the $\pi'$--$0'$
transition refers to the point in the $(T,T_K/\Delta)$-parameter space
where the global minimum of the free energy switches from  $\phi=\pi$ to  $\phi=0$ 
(triangles in Fig.~\ref{fig:phasediagram});
(iii) finally, the $0'$--$0$ transition corresponds to the conversion
of the local minimum at phase
$\phi=\pi$ into a maximum, that is, the slope of $I_s(\phi)$ changes
sign for phase $\pi$ (squares in Fig.~\ref{fig:phasediagram}).

The symmetrical BCS-Anderson model has also been
investigated by Siano and Egger~\cite{Siano04} using the quantum Monte Carlo
technique. However, Choi {\it et al.} 
point out that Ref.~\onlinecite{Siano04} does not consider  
the true low temperature limit and that the
scales, such as the Kondo temperature, differ exponentially from
the conventional definitions.\cite{Choi05,Siano05}

The zero temperature limit cannot be reached within the SNCA scheme. For this
purpose we have performed a perturbative RG analysis, analogous to
the poor man's scaling approach for the normal state~\cite{Anderson70}.
In the one-loop evaluation, the vertex from the impurity coupling
term
\begin{equation}
   \label{eq:kondo-coupling}
   H_{\rm Kondo}=\frac{J_K}{4}\sum\limits_{\sigma\sigma'\atop\alpha\alpha'}
                  \sum\limits_{\mathbf{kk'}}
         \crop{f}{\alpha}\vec{\tau}_{\alpha\alpha'}\deop{f}{\alpha'}\;
         \crop{c}{\mathbf{k}\sigma}\vec{\sigma}_{\sigma\sigma'}\deop{c}{\mathbf{k}\sigma'}
\end{equation}
generates particle-particle and particle-hole loops of conduction
and pseudofermion Green's functions. Here $J_K$ is the Kondo coupling
($J_K=V^2/|\epsilon_d|$ for $U\rightarrow\infty$) and $\tau^i$ ($\sigma^i$) are
the Pauli matrices in the impurity spin space (conduction electron
spin space).

While these diagrams
renormalize the Kondo coupling in the normal state, one-loop contributions with an
anomalous conduction electron propagator are to be included for the
superconducting state. Although the corresponding vertex is missing
in the bare Hamiltonian, the RG flow will generate the coupling which
is of the form
\begin{equation}
   \label{eq:anomalous-kondo-coupling}
   H_{g}=\sum\limits_{\sigma\sigma'\atop\alpha\alpha'}
                  \sum\limits_{\mathbf{k}} \frac{g_{ij}}{4}
         \crop{f}{\alpha}{\tau}^i_{\alpha\alpha'}\deop{f}{\alpha'}\;
         \crop{c}{\mathbf{k}\sigma}{\sigma}^j_{\sigma\sigma'}\crop{c}{\mathbf{-k}\sigma'}
\end{equation}
This coupling term will have the effect of cutting the RG flow for
small $T_K/\Delta$, that is, in the perturbative regime. Only the
coupling term with $g_{02}$ will be renormalized under the RG flow,
a consequence of  spin conservation and the symmetry of the order parameter.

With the initial conditions of
isotropic spin coupling and zero potential scattering term, the
following RG equations are obtained:~\cite{Sellier03}

\begin{subequations}
\label{eq:RGflow}
\begin{align}
   &\frac{d{J}}{d\ln D}=-\Re
         \left[\frac{D}{\sqrt{D^2-\Delta^2}}\right]
            \left({J}^2 -2\frac{\Delta}{D}{J}{g}\right)
                                          \\
   &\frac{d{g}}{d\ln D}=-\Re
         \left[\frac{D}{\sqrt{D^2-\Delta^2}}\right]
            \left(2\frac{\Delta}{D}{g}^2
                 +\frac{3}{2}\frac{\Delta}{D}{J}^2\right)
\end{align}
\end{subequations}
where $D$ is half the band width and the dimensionless couplings are defined as:
\begin{equation}
{J}=N_0 J_K \quad {\rm and} \quad {g}=-i N_0\, g_{02}\,.
\end{equation}
Here $g$ denotes the local coupling of pair fluctuations to the
impurity with the initial condition $g(D_0)=0$ for the bare band
cut-off $D_0$. For $\Delta\rightarrow 0$ one recovers the standard
poor man's scaling result.

\begin{figure}[b]
  \begin{center}
          \includegraphics[width=7.6cm]{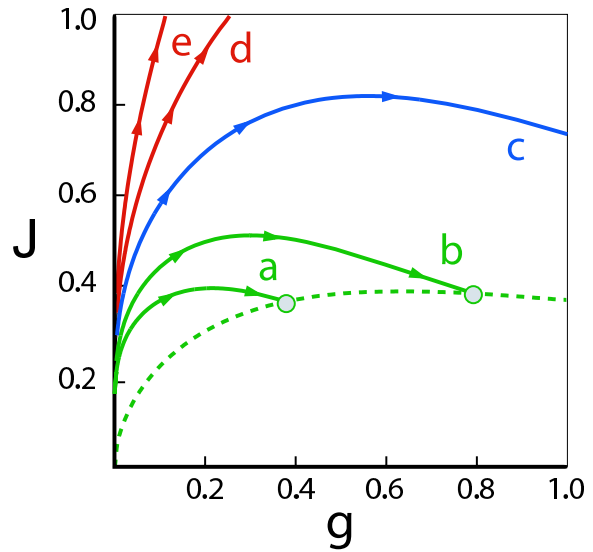}
    \caption{Scaling trajectories for various bare coupling values:
             (a) $T_K/\Delta=0.1$, (b) $T_K/\Delta=0.2$,  (c) $T_K/\Delta=0.5$,
             (d) $T_K/\Delta=1$, and (e) $T_K/\Delta=2$. The open
          circles represent fixed points, the dashed line is the fixed
          point line. The last fixed point in the perturbative regime
          is found for $T_K/\Delta\simeq 0.25$. For $T_K/\Delta\simeq
          0.65$ no downturn of the trajectory is observed. In the
          limit $T_K/\Delta\rightarrow\infty$ the trajectory follows
          the $J$-axis towards strong coupling.
}
    \label{fig:RGflow}
  \end{center}
\end{figure}

For weak coupling ($T_K/\Delta\lesssim 0.25$) the effective bandwidth
$D$ approaches $\Delta$ before $J$ or $g$ diverge; the square root in 
the scaling Eqs.~(\ref{eq:RGflow}) then vanishes and cuts off the RG flow.
The scaling trajectories in
the ${J}$--${g}$ plain flow towards a line of such fixed points.
The ground state is an unscreened spin. The strong coupling regime
is only accessible within a non-perturbative analysis but the tendency
of the trajectories to flow away from the fixed point line towards a
strong coupling fixed point with ${J}=\infty$ is already
observed in the present one-loop evaluation. The strong coupling fixed
point is approached for a bare coupling strength above a minimal $T_K/\Delta$ 
somewhere in between 0.25 and 0.65.

The solid fat line at the horizontal axis in
Fig.~\ref{fig:phasediagram} indicates the range where this quantum
phase transition ($\pi'$--$0'$ transition) is supposed to take
place. The range is certainly too wide in order to estimate the
low temperature extrapolation of the intermediate temperature
data. However the transition range is consistent with the SCNA
results. This does not apply for the extrapolated transition value
of Clerk and Ambegaokar~\cite{Clerk00b} which is already at
positive values of log$(T_K/\Delta)$. We do not well understand
the discrepancy to the result of Clerk and Ambegaokar; it may be
related to the way in which the zero temperature limit was
approached in Ref.~\onlinecite{Clerk00b}. However the NRG result
of Choi {\it et al.}~\cite{Choi04} (with transition at
$T_K/\Delta\simeq 0.42$, star in
Fig.~\ref{fig:phasediagram}) for the particle-hole symmetric model
is within  our estimate from the 1-loop RG. 
Yet this comparison should be taken with caution as the particle-hole
symmetric Anderson model also allows for local equal-time pair correlations
(of the $f$-particles) and has no potential scattering term, both in 
contrast to the infinite-$U$ case. The potential scattering term present in
the asymmetric model induces a characteristic shift of the Kondo resonance
relative to the Fermi energy,\cite{Kirchner04} 
 and thus is expected to influence the 
 $0$--$\pi$ transition as well.
The transition from the spin doublet to a singlet state of a
Kondo impurity in a bulk superconductor was calculated within NRG by
Satori {\it et al.}~\cite{Satori92}. They found $T_K/\Delta\simeq
0.3$ which is presented by the open triangle in
Fig.~\ref{fig:phasediagram}. This NRG result is within the range of
our 1-loop RG estimate and appears to be in agreement with the
finite-temperature SNCA data.

It should be obvious from the phase diagram that the transition is not only achieved
by a change of the coupling parameter $T_K/\Delta$ (through, e.g.,
gating the quantum dot) but also by a temperature variation, provided
that the (fixed) coupling is in an intermediate range.
\section{Zero Bias Conductance}
\label{sec:conductance}
The Josephson current directly probes the phase-sensitive anomalous
Green's functions (see Eq.~(\ref{eq:critical_current})). The 
conductance, however, is related to the quasiparticle current
and it is expressed through the imaginary parts of the
diagonal Green's functions: it has a finite, measurable value if the derivative of
the Fermi function is not exponentially small and if the bulk density
of states (i.e., $\Im G^A(\omega$)) and the impurity spectral function
($\Im\gf{\mathcal{G}}{d}{A}(\omega)$) are both finite in the same
frequency interval (see Eq.~(\ref{eq:conductance})). This implies that
the conductance vanishes exponentially for the zero temperature limit
as we restrict our considerations to $s$-wave superconductors.

\begin{figure}[t]
  \begin{center}
      \includegraphics[width=8.6cm]{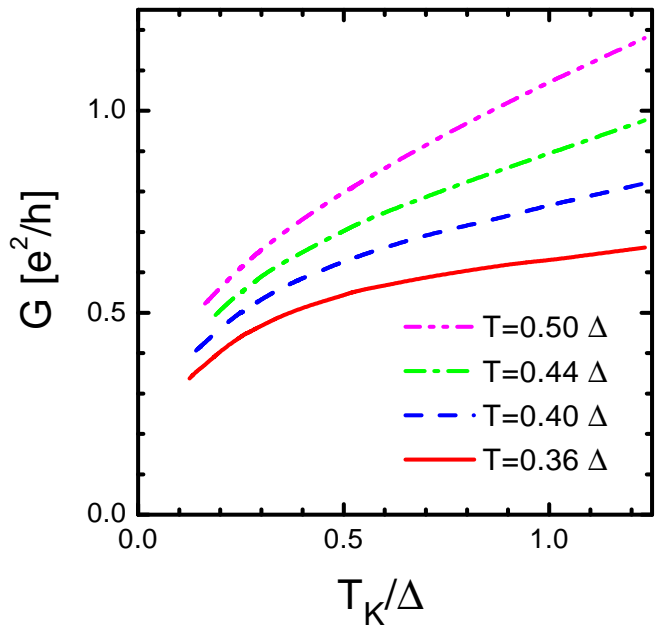}
    \caption{Zero bias conductance $G$ as a function of coupling strength
      $T_K/\Delta$ for fixed temperature.} 
    \label{fig:conductance_versus_TK}
  \end{center}
\end{figure}

The question arises if this conductance of the
quantum dot exhibits a signature of the $0$--$\pi$ transition at all. The
solution to this elementary question is not straightforward since the
conductance does not directly expose
the phase dependence of the superconducting states in the leads.
However, as the position of the subgap resonance moves from below through the
Fermi energy with increasing coupling, the ground state
transits into a singlet state through the Kondo screening of the impurity
spin. The enhanced screening, which is possible for strong coupling, 
not only modifies the subgap resonance but also the continuum through
increased pair breaking in this regime (cf.\ Fig.~\ref{fig:spdens}). 
Correspondingly, one may expect a feature in the temperature-dependent or 
coupling- ($T_K/\Delta$-) dependent conductance which signifies the 
transition.
This can be expected only if the temperature is not too small
(with respect to $\Delta$) so that $\partial f/\partial\omega$ in
Eq.~(\ref{eq:conductance}) is still sizable for frequencies with finite $\rho(\omega)$.
 
\begin{figure}[b]
 \begin{center}
      \includegraphics[width=8.5cm]{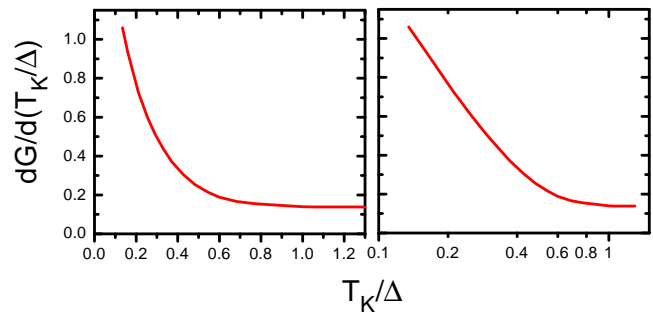}
    \caption{Derivative of the zero bias conductance, $G(T_K/\Delta)$,
       with respect to the coupling parameter  at
       $T/\Delta=0.36$. Left frame: linear scales; the
        $\pi'$--$0'$ transition is at approximately
        $T_K/\Delta\simeq 0.4$, for strong coupling $dG/d(T_K/\Delta)$ is
        constant. Right frame: coupling scale is in
      logarithmic presentation; for weak coupling $dG/d(T_K/\Delta)$
      diverges logarithmically.
              } 
    \label{fig:conductance_lin_log}
  \end{center}
\end{figure}

Consequently, the conductance $G$, Eq.~(\ref{eq:conductance}), 
represents an integral with respect to
the quasiparticle energy $\omega$ over an interval of the order of $T$. 
In order to exhibit characteristic features of the local spectral
density in the finite-$T$ conductance it is, therefore, 
suggestive to analyse the temperature derivative $dG/dT$.
Equivalently, one may consider the coupling parameter derivative, 
$dG/d(T_K/\Delta)$,
because $G$ is expected to be a universal function 
in terms of $T/T_K$ for fixed $\Delta$.
At finite $T$ and fixed $\Delta$,
both $dG/dT$ and $dG/d(T_K/\Delta)$ may be expected, in a rough first
estimate, to be essentially proportional to 
$\Im\gf{\mathcal{G}}{d}{A}(\omega = T)$ and 
$\Im\gf{\mathcal{G}}{d}{A}(\omega = T_K)$, respectively,
disregarding the energy dependence of the quasiparticle density
of states in the leads --- a more refined analyses certainly has to take
the detailed frequency structure of the integrand into account. 
Hence, in the weak coupling regime 
($\ln (\Delta /T_K) \gg 1$, $\ln (T /T_K) \gg 1$)     
we expect approximately that $dG/d(T_K/\Delta )\propto - \ln (T_K/\Delta)$,
while in the strong coupling region ($T_K/\Delta \gtrsim 1$, $T/\Delta < 1$),
$dG/d(T_K/\Delta )$ should be approximately $T$ and $T_K$ independent.
This expected linear behavior of $G$ for large coupling strength,
where Kondo screening is dominant, is associated with the formation of the 
Kondo resonance as it collects spectral weight and saturates at 
energies below $T_K$. 

For the numerical evaluation we focus on the
lowest temperature in Fig.~\ref{fig:conductance_versus_TK} 
($T/\Delta=0.36$, continuous line). The conductance
displays a constant slope in the $0$-phase 
($T_K/\Delta \gtrsim 0.4$), 
whereas in the $\pi$-phase ($T_K/\Delta \lesssim 0.4$)
the slope of $G$ tends to diverge for $T_K \to 0$.
This crossover is even more apparent
from the derivate of the conductance with respect to $T_K/\Delta$ (see
the left panel of Fig.~\ref{fig:conductance_lin_log} and the discussion above).  
Clear logarithmic behavior of the slope is observed
in the $\pi$-phase, where the numerical SNCA results are
especially controlled (see right panel of Fig.~\ref{fig:conductance_lin_log}).
The crossover between constant slope and logarithmic behavior
is found to be at a coupling strength which correponds to the $0$--$\pi$
transition. 
In fact, for $T/\Delta=0.36$, the $\pi'$--$0'$ transition is at approximately
$T_K/\Delta\simeq 0.4$. 
\begin{figure}[b]
  \begin{center}
      \includegraphics[width=7.5cm]{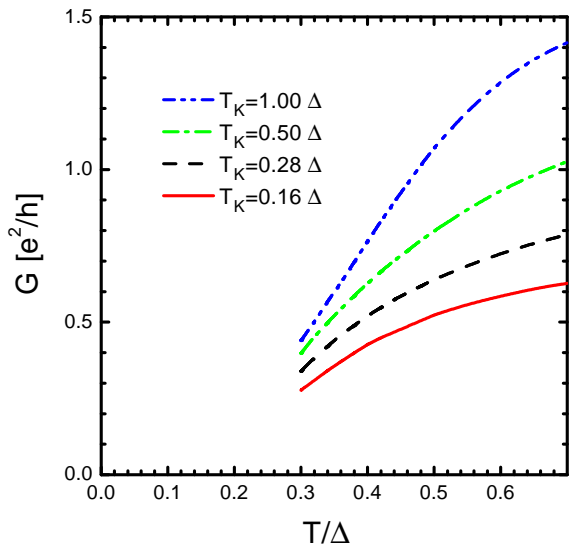}
    \caption{Zero bias conductance $G$ as a function of temperature
      $T/\Delta$ for fixed coupling strengths $T_K/\Delta$.
     }
    \label{fig:conductance_versus_T}
  \end{center}
\end{figure}

Finally, we explore the temperature dependence of the quantum dot conductance.
In Fig.~\ref{fig:conductance_versus_T} we present the SNCA results for
the zero bias conductance $G$ versus temperature $T$ at fixed values of $T_K/\Delta$. 
The exponential regime for very low temperatures is outside the
range where the SNCA is reliable. For the low temperature limit $\partial f/\partial\omega$
is exponentially small for $\omega>|\Delta|$ and should control the
temperature dependence of the conductance. However, with our 
data we are still in the regime with a wide derivative of the Fermi
function and a negative curvature of the
$G(T)$ lines. Fig.~\ref{fig:conductance_versus_T} suggests to assign a 
constant slope to the conductance in this intermediate temperature
range for the strong coupling regime. 
Such a linear behavior in $G(T)$, Eq.~(\ref{eq:conductance}), is
generated by an approximate compensation of the sharp spectral structures in $\rho(\omega)$
and in $\Im\gf{\mathcal{G}}{d}{A}(\omega)$ at the gap edge. 
The integration over the derivative of the Fermi function yields the
linear dependence for the intermediate temperature range --- although 
the non-leading contributions to the frequency dependence
above the gap edge may alter the temperature dependence. The linear
temperature behavior is consistent with the linear $T_K$ dependence of
$G$ which was discussed before. For the weak coupling regime one
should expect a $\ln T$ dependence of $dG/dT$. The temperature range
for the lowest curve in  Fig.~\ref{fig:conductance_versus_T} (with
$T_K/\Delta=0.16$) is too narrow to decide about a logarithmic
divergence of $dG/dT$ but the few data points are consistent with this assumption.

\section{Conclusion}
In conclusion, we have calculated the Josephson current as well as the 
linear response quasiparticle conductance for quantum dots in the Kondo
regime with superconducting leads. For finite temperatures, we have 
used a superconducting extension of the non-crossing approximation (SNCA),
while for $T=0$ the behavior was analyzed using a perturbative 
renormalization group treatment. In this way we mapped out the phase 
diagram of the $0-\pi$ transition of the Josephson current in the 
parameter space of temperature and Kondo coupling constant for the
first time in a systematical way. 
We stress that for temperatures not too far below $T_K$ the SNCA is 
expected to produce reliable, semiquantitative results.
The $T\to 0$ extrapolation of our results agrees well with
an NRG treatment of the problem at $T=0$,\cite{Choi04} considering that the latter
was done for the finite-$U$ symmetric Anderson model which, on one hand,
allows for equal-time Cooper pair formation on the dot, and, on the other
hand, does not have a potential scattering term --- both in contrast to 
our asymmetric, infinite-$U$ model. Considering finite $T$ will be 
essential for the analysis of experiments (see below).

Our results confirm that the Josephson current undergoes 
a succession of three transitions, $0-0'$, $0'-\pi'$, $\pi '-\pi$, 
separating four different Josephson equilibrium states, as the
Kondo temperature $T_K$ is reduced below the superconducting gap energy 
$\Delta$, or, alternatively, as the temperature $T$ is raised above $T_K$.
In going from a 0-type to a $\pi$-type junction, 
the four types of Josephson junctions are related to the successive
development of the minima of the free energy\cite{Siano04,Rozhkov99}.
By explicit calculations we could relate these phases to the characteristic
phase dependence of the Josephson current as well as to the position of the
Kondo-like subgap resonance above or below the Fermi energy. 
Moreover, we have also identified the signature of the $0$-$\pi$ transition 
in the quasiparticle linear response conductance $G$. Since the latter 
vanishes exponentially for $T\to 0$, a treatment at finite $T$ was essential
here. From our results, 
the $0$-junction regime appears to be characterized by a constant slope
of the conductance as a function of $T_K/\Delta$, while in the 
$\pi$-junction regime the slope diverges locarithmically. 
Note that at finite $T$ the $0$-$\pi$ transitions are continuous 
crossovers because of the finite width of the subgap resonance and develop
a discontinuous jump only for $T\to 0$ with vanishing width of the subgap 
resonance.\cite{Choi04} 
These relations may be relevant for identifying and analyzing 
the different phases in experiments like quantum dots, carbon nanotubes or 
other Kondo molecules coupled to superconducting leads. Gated devices
are supposed to control the level position in the quantum dot and,
correspondingly, the Kondo scale.

\begin{acknowledgments}
The authors are grateful to Peter W\"olfle for discussions and
support. G.S. and J.K.~acknowledge discussions with Stefan Kirchner.
This work is supported by 
BMBF 13N6918A (T.K, G.S.), DAAD D/03/36760 (T.K.), NSF-INT-0340536
(Y.S.B.), RFBR grant No. 05-02-17175 (Y.S.B.) and by
the Deutsche Forschungsgemeinschaft through CFN (G.S.), SFB~484 (T.K., G.S.)
and through grant No. KR1726/1 (J.K.).
\end{acknowledgments}
\appendix
\section{Superconducting NCA (SNCA)}
\label{ap:SNCAformulae}
In this appendix, we give a detailed derivation and discussion of the
formulae for the selfenergies, the auxiliary particle propagators 
and the local electron Green's function within the SNCA. The same 
approximation has been used by Clerk and Ambegaokar in 
Refs.\ \onlinecite{Clerk00b,Clerk00a}.

The BCS-Anderson Hamiltonian Eq.~(\ref{eq:bcsanderson}) obeys a 
local U(1) gauge symmetry with respect to simultaneous, time dependent 
transformations of the auxiliary particle fields, 
\begin{equation}
\label{eq:gauge}
\deop{f}{\sigma}\to {\rm e}^{i\varphi(t)}\deop{f}{\sigma}\ , \ \
\deop{b}{}\to {\rm e}^{i\varphi(t)}\deop{b}{}\ , \ \
\lambda \to \lambda - \frac{\partial \varphi}{\partial t} \,,
\end{equation}
which is intimately related to the conservation of the local charge $Q$,
and which, due to Elitzur's theorem, cannot be broken.\cite{Elitzur75}
The local gauge symmetry is implemented in the standard 
way,\cite{Kadanoff61} where a   
conserving, selfconsistent approximation is generated from a Luttinger-Ward 
functional via functional derivative with respect to 
the renormalized pseudoparticle propagators as well as the lead Green's 
functions.\cite{Kroha97,Kroha98}
In addition, the constraint $Q=1$ is enforced in any expectation 
value of a {\it physical} operator acting on the impurity state 
(more precisely: any operator which annihilates the $|Q=0\rangle$ state) 
by taking the limit $\lambda \to \infty$; e.g. for the 
physical $d$-electron Green's function,\cite{Coleman84,Costi96}
\begin{equation}
\label{eq:projection}
{\cal G}_{d\sigma}(t)= -i \lim _{\lambda\to\infty} \
\frac{\langle \deop{d}{\sigma}(t)\crop{d}{\sigma}(0)\
{\rm e}^{-\beta H - \beta\lambda (Q-1)} \rangle}
{\langle Q\ {\rm e}^{-\beta H - \beta\lambda (Q-1)}\rangle}\, ,
\end{equation}
with $\langle\, \dots \rangle$  the time-ordered, grand canonical
expectation value and $\beta = 1/T$. 

{\it Implications of the projection $Q=1$.} 
The constraint crucially influences the 
auxiliary particle dynamics and, in particular, prohibits
any anomalous contributions to the auxiliary particle propagators,
even in the case of superconducting leads. Consider, for example,
the Nambu pseudofermion propagator,
\begin{eqnarray}
\label{eq:gfinv}
{\mathbf F}_{\sigma}(\omega ) &=& 
-i \Theta (t) \langle \{ 
\left(
\begin{array}{c}
\deop{f}{\sigma}\\ \crop{f}{-\sigma}\\
\end{array}
\right) ,
\left(
\crop{f}{\sigma}\ \deop{f}{-\sigma}
\right)
\} 
\rangle\Biggr| _{\omega}\\
&&\hspace*{-1.7cm} =\left( 
\begin{array}{cccc}
\omega - \epsilon_d -\lambda -\Sigma _{\sigma}(\omega)& 
                  -\Sigma_{\sigma}^{anom}(\omega)\\
-\Sigma_{\sigma}^{anom\ *}(-\omega)                   &
\omega + \epsilon_d +\lambda +\Sigma _{-\sigma}(-\omega)\\
\end{array}
\right) ^{-1} , \nonumber
\end{eqnarray}
where $\Sigma_{\sigma}$ 
is the normal selfenergy, and the anomalous selfenergy
$\Sigma_{\sigma}^{anom}$ is assumed non-zero for the moment. Using the 
gauge $(\omega - \lambda) \to \omega$, performing the matrix inversion
in Eq.~(\ref{eq:gfinv}), and then taking the limit $\lambda\to\infty$
proves that all but the (11) element of the 
pseudo\-fermion propagator, $F_{\sigma} (\omega ) \equiv 
[{\mathbf F}_{\sigma}(\omega)]_{11}$, vanish. 
An analogous proof holds for the slave boson propagator. 
As a result, the (retarded) pseudofermion and slave boson propagators 
$F^R(\omega)$, $B^R(\omega)$, respectively, have only normal contributions,
\begin{subequations}
\label{eq:sncaprop}
\begin{align}
   \gf{F}{}{R}(\omega)&=\left(\omega-\ed-\lambda-\gf{\Sigma}{}{R}(\omega)\right)^{-1} \\
   \gf{B}{}{R}(\omega)&=\left(\omega-\lambda-\gf{\Pi}{}{R}(\omega)\right)^{-1}\,,
\end{align}
\end{subequations}
where $\gf{\Sigma}{}{R}(\omega)$, $\gf{\Pi}{}{R}(\omega)$ are the 
pseudofermion and slave boson selfenergies, respectively, and
the spin index has been suppressed in the absence of a magnetic field.

The evaluation of the Matsubara sum over a pseudoparticle frequency
shows that each closed pseudoparticle loop carries a 
fugacity factor ${\rm e}^{-\beta\lambda}$. Hence, upon the 
projection $\lambda \to \infty$ the Luttinger-Ward generating 
functional is comprised of diagrams which contain 
exactly one closed pseudoparticle loop: Its vanishing fugacity
factor is cancelled by the corresponding factor
${\rm e}^{-\beta\lambda}$ of the term 
${\langle Q\ {\rm e}^{-\beta H - \beta\lambda (Q-1)}\rangle}$ in the 
denominator of any physical expectation value 
[see Eq.~(\ref{eq:projection})].
\begin{figure}[t]
  \begin{center}
    \begin{tabular}{l l}
    $\gf{\Sigma}{}{}(i\omega)\;=$
    &\begin{minipage}[b]{2.21cm}
      \includegraphics[width=2.21cm]{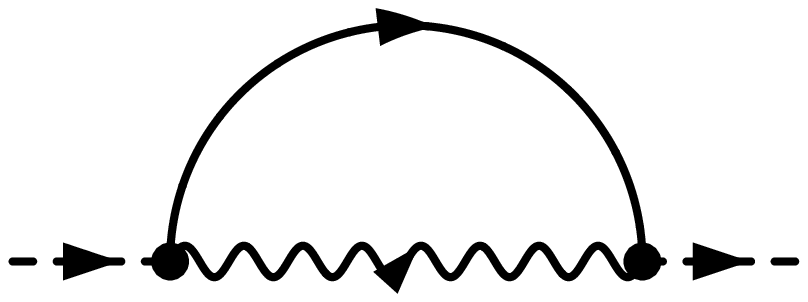}
    \end{minipage}
    $+$
    \begin{minipage}[b]{3.47cm}
      \includegraphics[width=3.47cm]{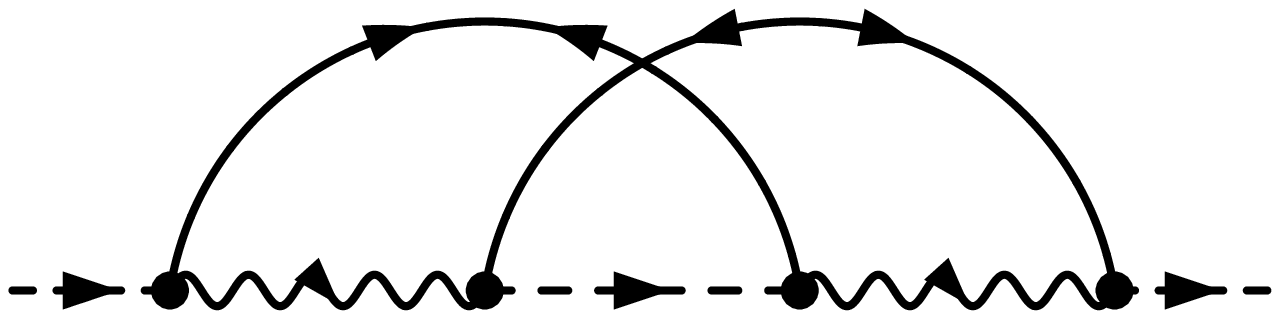}
    \end{minipage}\\[0.5cm]
    $\gf{\Pi}{}{}(i\nu)\;\;=$
    &\begin{minipage}[b]{2.22cm}
      \includegraphics[width=2.22cm]{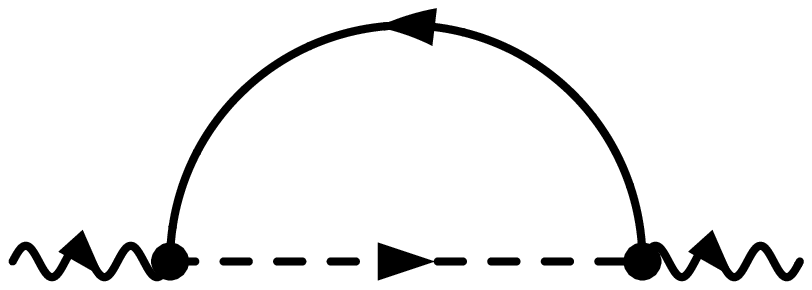}
    \end{minipage}
    $+$
    \begin{minipage}[b]{3.47cm}
      \includegraphics[width=3.47cm]{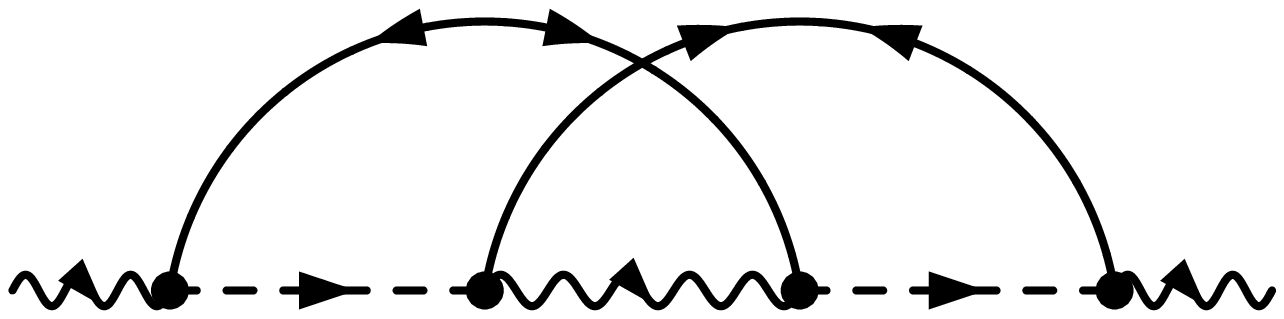}
    \end{minipage}
   \end{tabular}
   \caption{Pseudofermion selfenergy $\gf{\Sigma}{}{}(i\omega)$
            and slave boson selfenergy $\gf{\Pi}{}{}(i\nu)$}
   \label{fig:SelfenergiesSNCA}
  \end{center}
\end{figure}

{\it Definition of the SNCA.}
To define a conserving approximation that describes
coherent Cooper pair transmission through a Kondo
quantum dot, the generating functional shown 
diagrammatically in Fig.\ \ref{fig:GenFuncSNCA} has been chosen. 
The first diagram of Fig.\ \ref{fig:GenFuncSNCA} represents 
the conventional non-crossing approximation and describes 
normal quasiparticle transmission.
The second diagram of Fig.\ \ref{fig:GenFuncSNCA} contains 
two crossing anomalous superconducting Green's functions in the leads.
It is the simplest (i.e. lowest order in the hybridization $V_a$)
contribution to incorporate coherent Cooper pair tunneling in the
Luttinger-Ward functional. Therefore, the corresponding approximation has  
been termed ``superconducting non-crossing approximation'' (SNCA).
The resulting auxiliary particle selfenergies 
$\Sigma$, $\Pi$, are shown in Fig.\ \ref{fig:SelfenergiesSNCA}
and read,
\begin{subequations}
\label{eq:sncaself}
\begin{alignat}{2}
   &\gf{\Sigma}{}{R}(\omega)\,&=&\sum\limits_a\frac{\Gamma_a}{\pi}\integral{\epsilon}{}{}
        f(\epsilon)\,\rho_a(-\epsilon)\,\gf{B}{}{R}(\epsilon)\notag\\
   &&-&\sum\limits_{aa^\prime}\frac{\Gamma_a\Gamma_{a^\prime}}{\pi^2}
          \integral{\epsilon}{}{}f(\epsilon)\,g_a^*(\epsilon)
          \integral{\epsilon^\prime}{}{}f(\epsilon^\prime)\,g_{a^\prime}(\epsilon^\prime)\,\notag\\
   &&\times&\,\gf{B}{}{R}(\omega+\epsilon)\,\gf{B}{}{R}(\omega+\epsilon^\prime)\,
          \gf{F}{}{R}(\omega+\epsilon+\epsilon^\prime)\,,\\
   &\gf{\Pi}{}{R}(\omega)\,&=&\sum\limits_a\frac{N\,\Gamma_a}{\pi}\integral{\epsilon}{}{}
        f(\epsilon)\,\rho_a(\epsilon)\,\gf{F}{}{R}(\epsilon)\notag\\
   &&+&\sum\limits_{a^\prime}\frac{N\,\Gamma_a\Gamma_{a^\prime}}{\pi^2}
          \integral{\epsilon}{}{}f(\epsilon)\,g_a(\epsilon)
          \integral{\epsilon^\prime}{}{}f(\epsilon^\prime)\,g_{a^\prime}^*(\epsilon^\prime)\,\notag\\
   &&\times&\,\gf{F}{}{R}(\omega+\epsilon)\,\gf{F}{}{R}(\omega+\epsilon^\prime)\,
          \gf{B}{}{R}(\omega+\epsilon+\epsilon^\prime)\,,
\end{alignat}
\end{subequations}
where $f(\epsilon)$ is the Fermi distribution function and $N=2$ is the spin
degeneracy. As seen from Fig.\ \ref{fig:SelfenergiesSNCA}, the SNCA
incorporates exactly two coherent Andreev transmission terms,
one where a pseudofermion disappears from the dot and forms a Cooper pair
in the superconducting lead, leaving an additional, virtual
pseudofermion hole with opposite spin behind, 
and one describing the inverse process.  
Incoherent, sequential Andreev processes are included in the propagators 
to infinite order via selfconsistency. 
The set of equations (\ref{eq:sncaself}) can be further simplified
in extracting the explicit phase-dependence of the conduction electron
functions, i.e.
\begin{align*}
   &\sum\limits_a\Gamma_a\rho_a(\epsilon)=\Gamma\rho(\epsilon)\,,\\
   &\sum\limits_{aa^\prime}\Gamma_a\Gamma_{a^\prime}
       g_a(\epsilon)g_{a^\prime}^*(\epsilon^\prime)
       =\Gamma^2\cos^2\left(\frac{\phi}{2}\right)|g(\epsilon)||g(\epsilon^\prime)|\,.
\end{align*}
Here, we have defined
the phase difference $\phi=\phi_L-\phi_R$ and, and assumed symmetrical 
superconductors in the leads, $|g_L(\epsilon)|=|g_R(\epsilon)|
\equiv |g(\epsilon)|$, $\rho_L(\epsilon)=\rho_R(\epsilon)\equiv\rho(\epsilon)$.
By means of these relations it is seen immediately
that only the terms with anomalous conduction electron
propagators contribute to the phase dependence of the
selfenergies in Eqs.~(\ref{eq:sncaself}).\par
\begin{figure}[b]
  \begin{center}
    $\gf{\mathcal{G}}{d}{}(i\omega)\;=\;$
    \begin{minipage}[c]{1.46cm}
      \includegraphics[width=1.46cm]{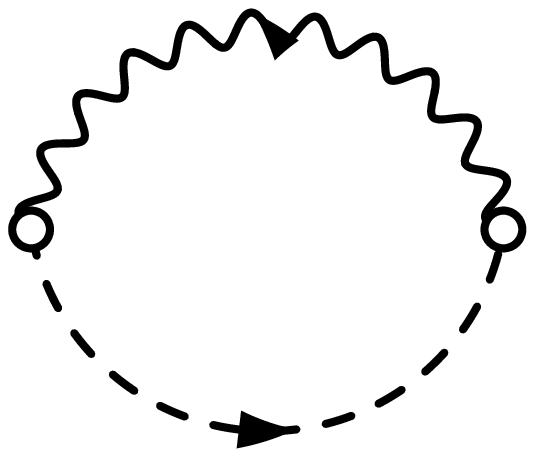}
    \end{minipage}\;\;,
    $\quad\gf{\mathcal{F}}{d}{}(i\omega)\;=\;$
    \begin{minipage}[c]{1.90cm}
      \includegraphics[width=1.90cm]{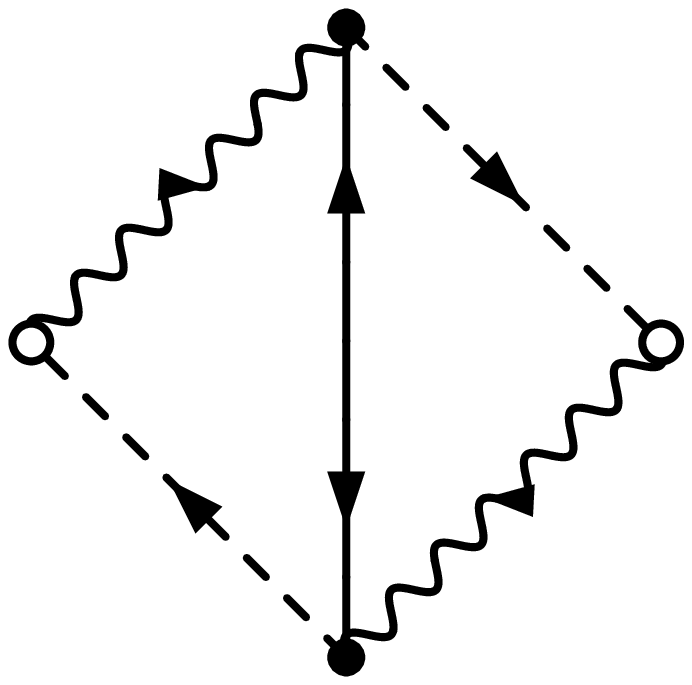}
    \end{minipage}\\
    \caption{Diagonal element $\gf{\mathcal{G}}{d}{}(i\omega)$
             and off-diagonal element $\gf{\mathcal{F}}{d}{}(i\omega)$ of the
             impurity Green's function.}
    \label{fig:TmatSNCA}
  \end{center}
\end{figure}
The equations for the normal and for the anomalous, physical dot electron
Green's function are derived analogously by functional derivative of 
the generating functional with respect to the corresponding 
lead electron propagators. This yields (see Fig.\ \ref{fig:TmatSNCA}),
\begin{subequations}
\label{eq:sncadgreens}
\begin{alignat}{2}
    \Im&\gf{\mathcal{G}}{d}{R}(\omega)&=&-\!\int\!\frac{d\epsilon}{\pi}
        \, e^{-\beta\epsilon}\,
        \Im\gf{F}{}{R}(\epsilon+\omega)\,\Im\gf{B}{}{R}(\epsilon)\,,\\
    \Im&\gf{\mathcal{F}}{d}{R}(\omega)&=&-\Gamma\cos\left(\frac{\phi}{2}\right)
        \!\int\!\frac{d\epsilon}{\pi}
        \, e^{-\beta\epsilon} 
        \!\int\!\frac{d\epsilon^\prime}{\pi}
        f(\epsilon^\prime)\,|g(\epsilon^\prime)|\:\notag\\
        &&\times&\Im\left[\gf{F}{}{R}(\epsilon+\omega)\,
                 \gf{B}{}{R}(\epsilon+\epsilon^\prime+\omega)\right]\notag\\
        &&\times&\Im\left[\gf{F}{}{R}(\epsilon+\epsilon^\prime)\,
                 \gf{B}{}{R}(\epsilon)\right]\,.
\end{alignat}
\end{subequations}
Without loss of generality we have set $\phi_L+\phi_R=0$.
Note that, although the local U(1) gauge symmetry on the dot prevents
anomalous contributions to the auxiliary particle propagators,
anomalous physical electron Green's functions on the dot do exist.
Physically this means that temporally retarded Cooper pairs 
on the quantum dot are indeed induced by the proximity effect, even 
though the formation of equal-time Cooper pairs is completely suppressed
by the local Coulomb repulsion.

{\it Numerical evaluation of the SNCA.}
The Eqs.~(\ref{eq:sncaprop}) and (\ref{eq:sncaself}) form a closed
set of non-linear integral equations for the auxiliary particle 
propagators which is solved numerically by iteration. 
The physical dot electron Green's functions, which determine 
the Josephson as well as the quasiparticle current (see Appendix B),
are then computed using Eqs.~(\ref{eq:sncadgreens}). 
Note, however, that the Boltzmann factors in Eqs.~(\ref{eq:sncadgreens}) 
strongly diverge for negative
frequencies. Although this divergence is compensated by the threshold
behavior of the pseudoparticle propagators, a numerical evaluation
necessitates a re-formulation in terms of a new set of functions,
$\Im\gf{\tilde{F}}{}{}(\omega)$, $\Im\gf{\tilde{B}}{}{}(\omega)$. 
We define them via the relations
\begin{align*}
   \Im\gf{F}{}{R}(\omega)=f(-\omega)\,\Im\gf{\tilde{F}}{}{}(\omega)\;,\\
   \Im\gf{B}{}{R}(\omega)=f(-\omega)\,\Im\gf{\tilde{B}}{}{}(\omega)\;.
\end{align*}
Since the Boltzmann factors appear precisely in conjunction with the
integrals along the branch cuts of the auxiliary particle Green's
functions, it is possible to absorb all these exponentially diverging 
factors in $\Im\gf{\tilde{F}}{}{}(\omega)$, $\Im\gf{\tilde{B}}{}{}(\omega)$
by observing ${\rm e}^{-\beta\omega}f(-\omega)=f(\omega)$.
Details of this method as well as of the efficient treatment
of the projection $\lambda \to \infty$ can be found in Ref.\ 
\onlinecite{Costi96}.
\section{Josephson Current Through an Interacting Region}
\label{ap:joscurrent}
A formula for the supercurrent through an interacting
region is derived. We proceed along the
line of references~\onlinecite{Meir92, Clerk00b}. The charge
current through the system can be expressed by means of
the time derivative of the electron numbers $N_{L/R}$ in
the left and right lead, i.e.
\begin{align}
  I_{L/R}=\mp\langle\dot{N}_{L/R}(t)\rangle\;,
\end{align}
Note that the relation $I_L=I_R$ holds. The time derivative
is calculated using the Heisenberg equation of motion, yielding the
expression
\begin{align}
  I_a=\eta_a\frac{i\,e}{\hbar}\sum\limits_{\mathbf{k}\sigma}
            V_a\left(
                    \langle\crop{d}{\sigma}\deop{c}{\mathbf{k}\sigma,a}\rangle
                   -\mathsf{h.c.}\right)
\end{align}
with $\eta_{L/R}=\mp 1$. Using the definition for the lesser
function $\gf{G}{\mathbf{k}\sigma a,\sigma^\prime}{<}(t,t^\prime)=
         i\langle\crop{d}{\sigma^\prime}(t^\prime)
                 c_{\mathbf{k}\sigma,a}(t)\rangle$,
$I_a$ may be rewritten as
\begin{align}
   I_a=\eta_a\frac{2e}{\hbar}
            \sum\limits_{\mathbf{k}\sigma}
            V_a\Re\left[\gf{G}{\mathbf{k}\sigma a,\sigma}{<}(0)\right]\,.
\end{align}
\begin{figure}[t]
    \vskip 0.5cm
    \begin{minipage}[t]{3.03cm}
      \includegraphics[width=3.1cm]{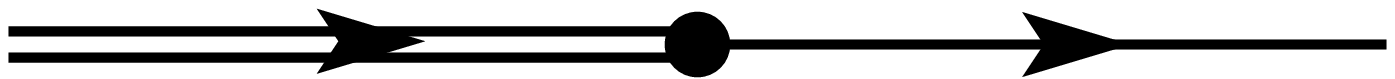}
    \end{minipage}
    \phantom{aaaaa}
    \begin{minipage}[b]{3.03cm}
      \includegraphics[width=3.1cm]{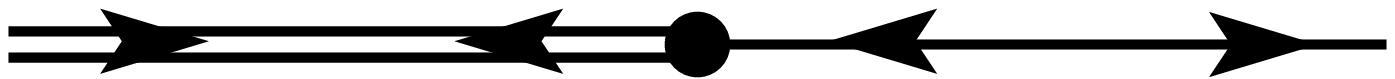}
    \end{minipage}
   \caption{Diagrams for $\gf{G}{\mathbf{k}\sigma a,\sigma}{<}$;
            The double line represents the dot electron Green's function,
            the single line the conduction electron green's function.
            The left diagram contributes to the normal current, the second
            diagram contributes to the supercurrent, Eq.~(\ref{eq:supercurrent})}
   \label{fig:glessdiagrams}
\end{figure}
The diagrams contributing to $\gf{G}{\mathbf{k}\sigma a,\sigma}{<}$
are shown in figure \ref{fig:glessdiagrams}. In the following, the
system is assumed to be in equilibrium, i.e. $e\mathrm{V}=0$. In
this limit the normal current vanishes, and charge is transfered
through the system only via the supercurrent $I_s$, with
\begin{align}
   \label{eq:supercurrent}
   I_{s,a}=\eta_a\frac{2e}{\hbar}
            \sum\limits_{\mathbf{k}\sigma}V_a^2
            \int\!\frac{d\omega}{2\pi}
            \Re &\left[\gf{\mathcal{F}}{d}{\dagger,A}(\omega)
                     \gf{F}{\mathbf{k} a}{A}(\omega)\notag\right.\\
                    &-\left.\gf{\mathcal{F}}{d}{\dagger,R}(\omega)
                     \gf{F}{\mathbf{k} a}{R}(\omega)\right]\,.
\end{align}
To derive the above equation we have used the relations
$\gf{G}{}{}(\omega)=\gf{G}{}{<}(\omega)-\gf{G}{}{A}(\omega)$ and
$\bar{\gf{G}{}{}}(\omega)=\gf{G}{}{R}(\omega)-\gf{G}{}{<}(\omega)$ for the
time ordered and anti-time ordered Green's functions,
respectively, and
$\gf{G}{}{<}(\omega)=f(\omega)\left(\gf{G}{}{A}(\omega)-\gf{G}{}{R}(\omega)\right)$
for the lesser functions. For convenience we extract the explicit phase
dependence from the off-diagonal Green's functions,
\begin{align*}
   &\gf{\mathcal{F}}{d}{R\, \dagger}(\omega)=\cos\left(\frac{\phi}{2}\right)
                                          \gf{\bar{\mathcal{F}}}{d}{R\, \dagger}(\omega)\,,\\
   &\gf{F}{a}{R}(\omega)=e^{i\phi_a}\gf{\bar{F}}{}{R}(\omega)\,,
\end{align*}
where we have introduced the retarded, local conduction electron Green's function
$\gf{F}{a}{R}(\omega)=\sum_\mathbf{k}\gf{F}{\mathbf{k} a}{R}(\omega)$.
Moreover, without loss of generality we have set $\phi_L+\phi_R=0$.
Finally, using local charge conservation in the stationary case, 
$I_{s,L}=-I_{s,R}\equiv I_{s}$, we obtain the following formula
for the Josephson current\cite{Clerk00b}
\begin{align}
  I_s=\frac{2e}{h}\frac{\Gamma}{\pi N_0}\sin(\phi)&\integral{\omega}{}{}\,f(\omega)\notag\\
            &\times\Im\left[\gf{\bar{\mathcal{F}}}{d}{R\, \dagger}(\omega)
                     \gf{\bar{F}}{}{R}(\omega)\right]\,,
\end{align}
with $\Gamma=\Gamma_L+\Gamma_R$.

\end{document}